# The speciation of Australopithecus and Paranthropus was caused by introgression from the Gorilla lineage

Johan Nygren
2018

ABSTRACT: The discovery of *Paranthropus deyiremeda* in 3.3–3.5 million year old fossil sites in Afar (Haile-Selassie, 2015), together with 30% of the gorilla genome showing lineage sorting between humans and chimpanzees (Scally, 2012), and a NUMT ("nuclear mitochondrial DNA segment") that is shared by both gorillas, humans and chimpanzees, and that dates back to 6 million years ago (Popadin, 2017), is conclusive evidence that introgression from the gorilla lineage caused the speciation of both the *Australopithecus* lineage and the *Paranthropus* lineage, providing a lens into the gorilla-like features within *Paranthropus*, as well as traits within *Homo* that originate from the gorilla branch, such as a high opposable thumb index (Almécija, 2015), an adducted great toe (Tocheri, 2011; McHenry, 2006), and large deposits of subcutaneous fat.

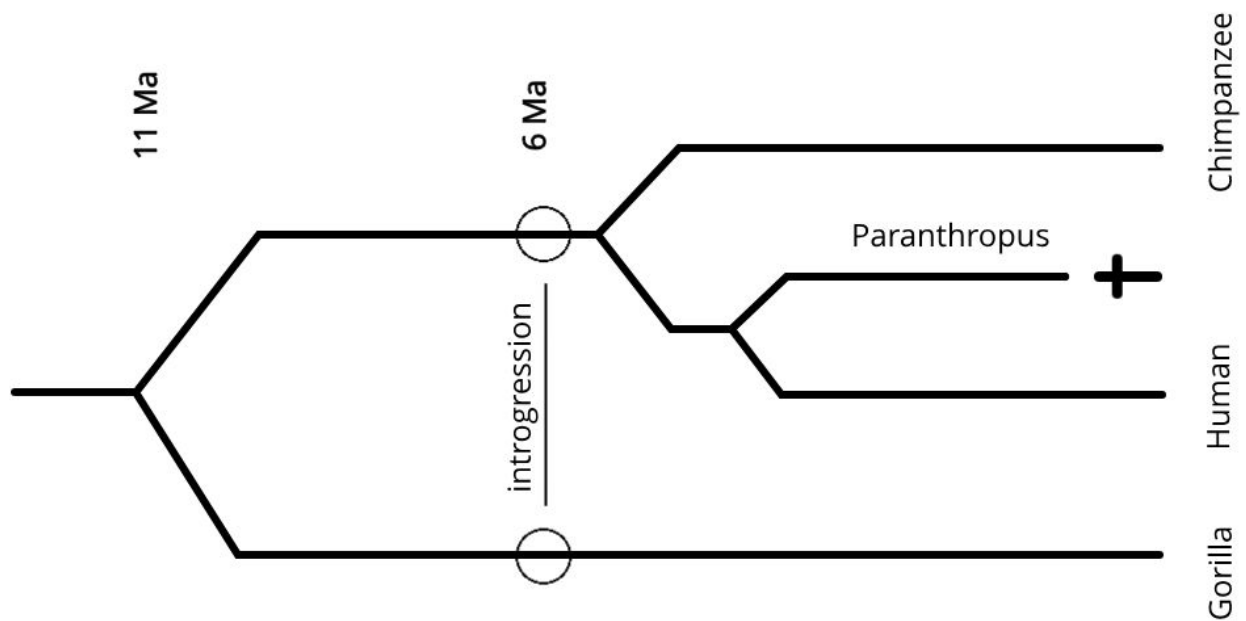

Fig. 1. Phylogenetic tree showing how introgression caused the speciation of humans. This introgression speciation model predicts an early split for *Paranthropus* and *Australopithecus*, increasingly shown in the fossil record (Haile-Selassie, 2015, 2016; Wood, 2016), and also shows that the evolution of genes that ended up in *Australopithecus*, and therefore in extant humans, as well as in *Paranthropus*, can and should be traced along the gorilla lineage as well.

Corresponding author: Johan Nygren, johanngrn@gmail.com, unaffiliated

## The introgression speciation model and a revised phylogenetic tree

The origin of our species has increasingly been discovered over the past century, through Darwin to the discovery of DNA and the double-helix, to Lucy and fossils of *Australopithecines* that originate from the Afar region in Ethiopia from the Pliocene. The genetic data from Scally in 2012, and Popadin in 2017, now provides conclusive evidence for how *Paranthropus* and *Australopithecus*, as two separate lineages, both speciated as a result of introgression from the gorilla lineage (Fig. 1). That introgression fills in the "missing link" and shows how the origin of our species is not one single continuous lineage, but the hybridization of both the gorilla lineage and the common ancestor of humans and chimpanzees, an event that occurred in the late Miocene. The introgression speciation model (Fig. 1) predicts an early split between the *Paranthropus* lineage and the *Australopithecus* lineage.

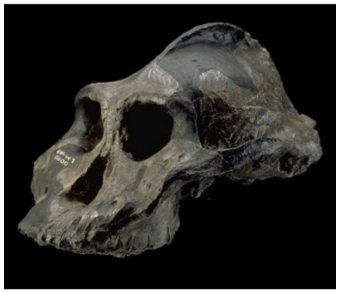 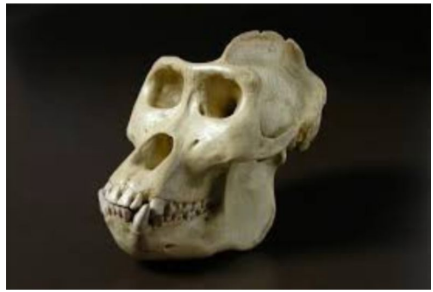 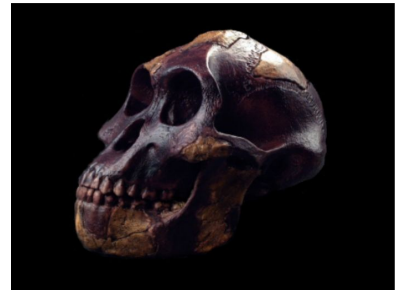

*Paranthropus aethiopicus*  *Gorilla gorilla*  *Australopithecus afarensis*
2.8-2.3 Ma                                    3.7-2.9 Ma

Fig2. Introgression from *Gorilla* caused the speciation of both *Australopithecus* and *Paranthropus*, and means that traits that have evolved independently in the gorilla lineage were transferred into the hybrid lineages. *Paranthropus* are often described as "gorilla-like", they have sagittal crests which suggest strong muscles of mastication, and broad, grinding herbivorous teeth, that led to the name "nutcracker man" for *Paranthropus boisei* who lived between 2.4–1.4 Ma.

## The Burtele foot (BRT-VP-2/73) and Au. deyiremeda, a Paranthropus?

With conclusive evidence of an introgression speciation model, it is apparent that the *Paranthropus* and *Australopithecus* lineages both speciated as a result of introgression from *Gorilla* (Fig. 2), and that the two lineages separated as early as 5-6 Ma, adapting to separate niches. The discovery of 3.2-3.5 million year old hominin fossils that show divergent evolution from *Au. afarensis* from the same time period (Haile-Selassie, 2012, 2015), featuring an abductable great toe (Fig. 3) instead of the human-like hallux of *Au. afarensis,* a human-like transverse arch that stiffens the foot (Haile-Selassie, 2012), instead of the transitional arch of *Au. afarensis* that is in-between *Homo* and *Pan*, and jaws and teeth that shares characteristics with *Paranthropus* and *Homo* (Haile-Selassie, 2015) suggested the classification of a new species *Australopithecus deyiremeda*, meaning "close relative" in the local Afar language.

The introgression speciation models shows that *Au. deyiremeda* is better classified as a *Paranthropus*, *P. deyiremeda*, and that an early split between *Paranthropus* and *Australopithecus* is the reason there were two separate lineages of hominins during the Pliocene (Haile-Selassie, 2015, 2016; Wood, 2016), clearly distinguishable by their locomotor adaptation and diet.

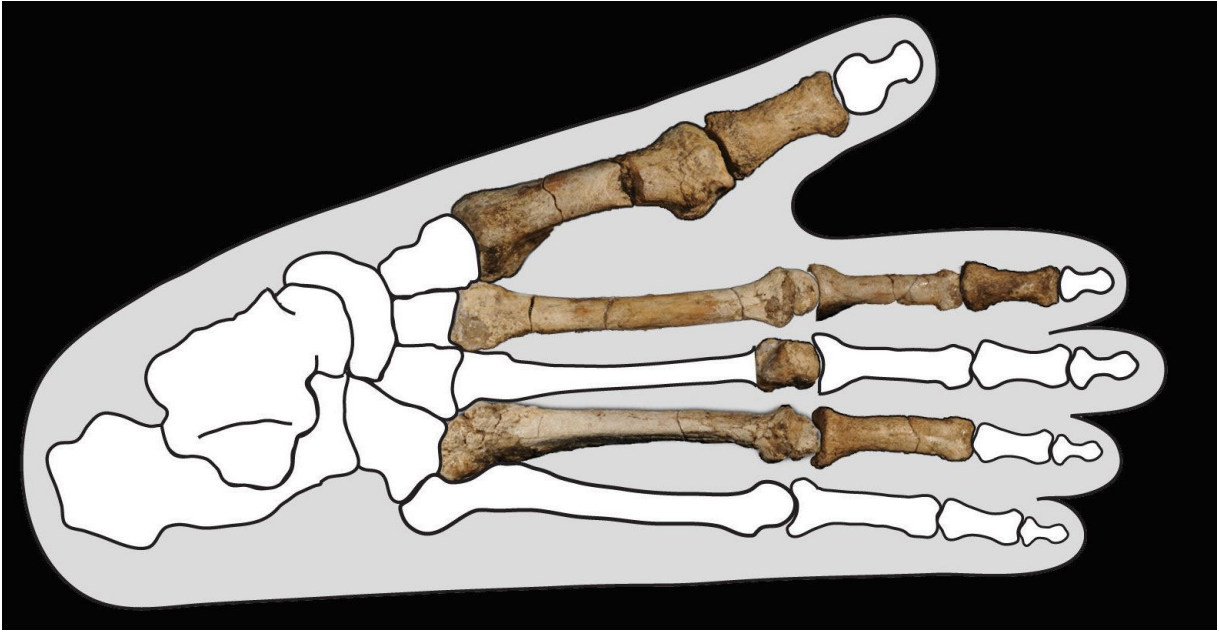

Fig 3. The Burtele foot, BRT-VP-2/73, found in 2009 (Haile-Selassie, 2015) in Burtele at Woranso-Mille, Afar, tentatively assigned *Au. deyiremeda* (Haile-Selassie, 2015), contemporaneous with *Au. afarensis*, shows distinct locomotor adaptation as it retains a grasping hallux, in contrast to the human-like foot that had developed in *Australopithecus afarensis*. The conclusive evidence of an introgression speciation model suggests that *deyiremeda* is better classified as *Paranthropus deyiremeda*, and that it provides a fossil record that support an early split for *Paranthropus* and *Australopithecus*, two lineages that adapted for separate niches.

## The classification of Paranthropus deyiremeda, an overview

The classification of *P. deyiremeda* within this thesis has its foundation within the genetic data for an introgression speciation model (Fig. 1), originally the Gorilla Genome Project (Scally, 2012), which shows without a doubt that there was introgression of 30% of the Gorilla genome into the human-chimpanzee lineage (the common ancestor of *Pan*/*Homo*. ) The introgression event, which can be dated from the NUMT on chromosome 5 to 4.5 Myr after the *Gorilla/Pan-Homo* split, to around 6 Ma, (Popadin, 2017), which is the time of the *Pan-Homo* split, is a boundary for the earliest split between *Paranthropus* and *Australopithecus*. Traits within *Paranthropus* that resemble *Gorilla*, such as the sagittal crest, are more parsimonious as a result of the introgression event rather than convergent evolution, and lineage sorting similar to the 30% of the Gorilla genome that displays lineage sorting with *Pan* and *Homo* (Scally, 2012), which supports the hypothesis of *Paranthropus* as a lineage that also speciated from the introgression.

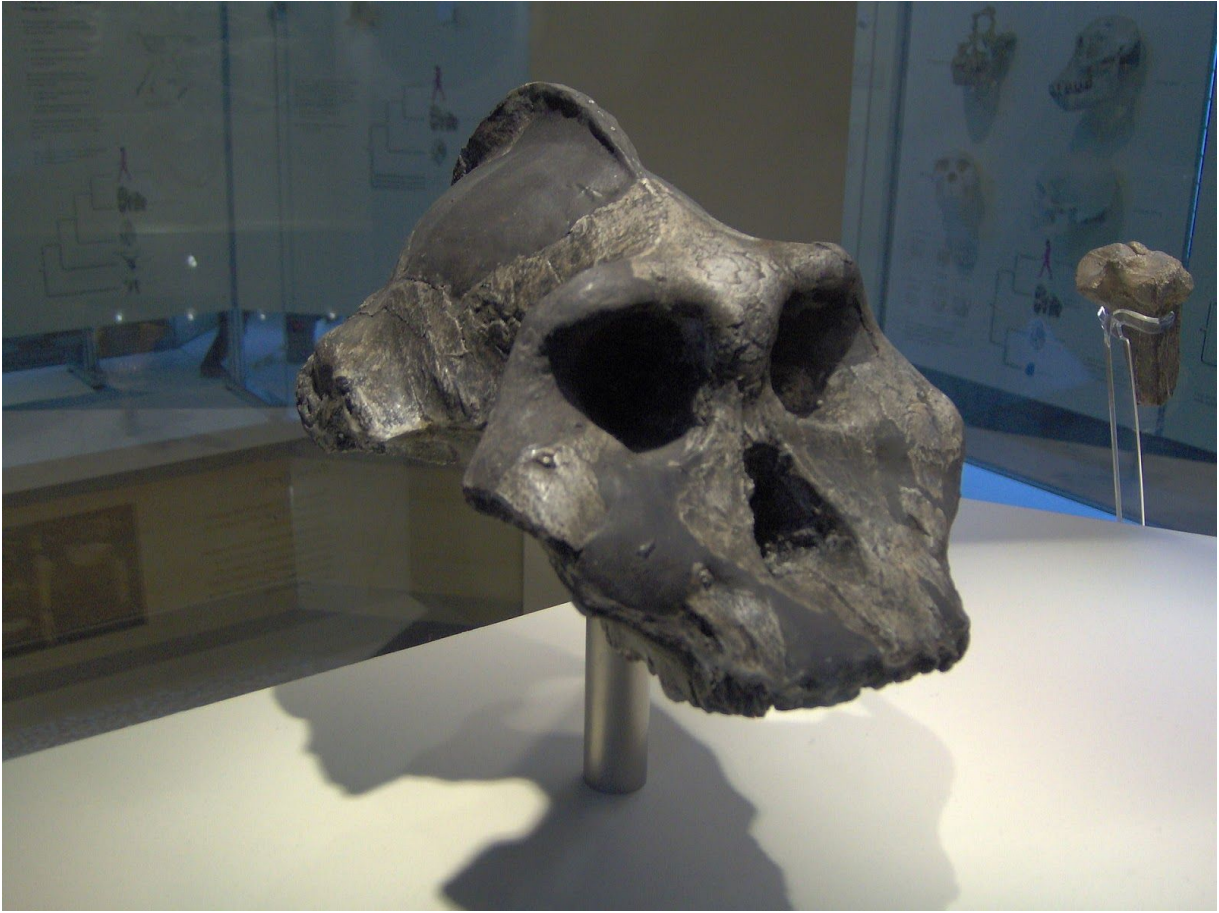

Fig. 4. *Paranthropus aethiopicus*, 2.8-2.3 Ma, with gorilla-like sagittal cranial crests as an attachment for strong muscles of mastication, a dietary adaptation. The genetic proof of an introgression event at the time of the *Pan-Homo* spit shows that the most parsimonious origin for those features within *Paranthropus* was lineage sorting from the introgression event, originating in *Gorilla*, rather than convergent evolution. Image from public domain (CC BY-SA 3.0).

The foot stiffness in *Paranthropus deyiremeda* (Haile-Selassie, 2012) is not a preserved character, it is a derived character that is absent in the *Au. afarensis* lineage as well as in *Pan* and *Gorilla*, and that exists together with an abducted great toe, and is contemporary with an adducted (human-like) hallux as a derived feature in *Au. afarensis*, undisputable data for that *deyiremeda* is a separate lineage that had adapted for a separate niche, which is also what justified its original classification as a "close relative" (Haile-Selassie, 2015). The denthognathic features that are similar to *Paranthropus* (Haile-Selassie, 2015) suggest similar dietary adaptations, and within the hypothesis of introgression as a cause of speciation, the most parsimonious explanation is lineage sorting from the introgression event, with adaptations for browsing such as large muscles of mastication that were co-opted for grazing. (Cerling, 2017)

## Introgression from Gorilla seen in lineage sorting between humans and chimpanzees

The past two decades have seen rapid advances in whole genome sequencing, with the Human Genome Project completed in 2003 (Venter, 2003), the chimpanzee genome two years later (Waterson, 2005) and the Gorilla genome in the following decade (Scally, 2012). The whole genome sequencing of *Gorilla* gave the first clues to an introgression speciation model, and showed that 30% of the gorilla genome exhibits lineage sorting with the human genome and chimpanzee genome (Scally, 2012), 15% of the gorilla genome is closer to *Homo* than to *Pan*, and another 15% closer to *Pan* than to *Homo*, a result of gene transfer from *Gorilla* in the introgression event that also transferred the NUMT on chromosome 5 (Popadin, 2017) to all three lineages.

## The NUMT on chromosome 5 as conclusive evidence of an introgression speciation model

Mitochondria is inherited separately from nuclear DNA, and fragments of mitochondrial DNA are known to get inserted into nuclear DNA to form NUMTs, i.e. nuclear pseudogenes of the mtDNA. There is a NUMT sequence on chromosome 5 shared by gorillas, chimpanzees, and humans, shown from mutation rates to date back to 6 million years ago, a result of interbreeding between lineages that had diverged as much as ~4.5Myr prior to the interbreeding event (Popadin, 2017) (Fig. 5), which fits with the *Gorilla*/*Pan-Homo* split at 11Ma. The NUMT on chromosome 5 shares affinities with the gorilla lineage mtDNA (Popadin, 2017) suggesting that it originates from the gorilla lineage.

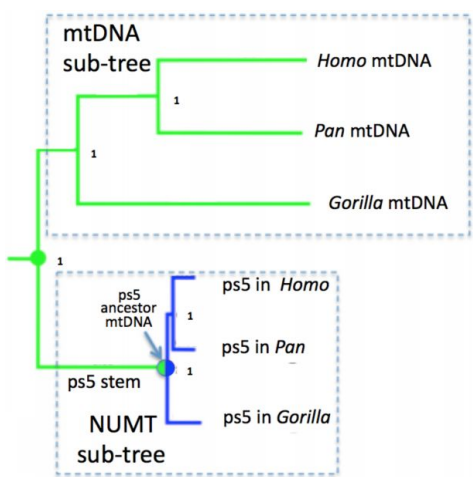

Fig 5. NUMT on chromosome 5 shared by *Gorilla*, *Pan* and *Homo*, that had diverged ~4.5Myr prior to the introgression event, which fits the estimate for when the *Gorilla* lineage split from the common ancestor of *Pan* and *Homo* around 11 million years ago. Image from Popadin et al, 2017.

# The introgression speciation model and the Pthirus host switch 3.3 million years ago

The evolutionary history of anthropoid primate lice includes a host switch from the gorilla lineage to *Homo* around 3.3 million years ago (Reed, 2007) (Fig. 6). The introgression speciation model supports the idea of Paranthropus as an intermediary in the host switch, and that *Paranthropus*, which shares ancestral features such as a cone shaped rib cage, abducted great toes, and similar diet as *Gorilla* (Cerling, 2011), may have continued to mate with both *Gorilla* and *Australopithecus* up to the Pthirus split 3.3 Ma, while *Australopithecus* were sexually isolated from *Gorilla*. (Fig. 7)

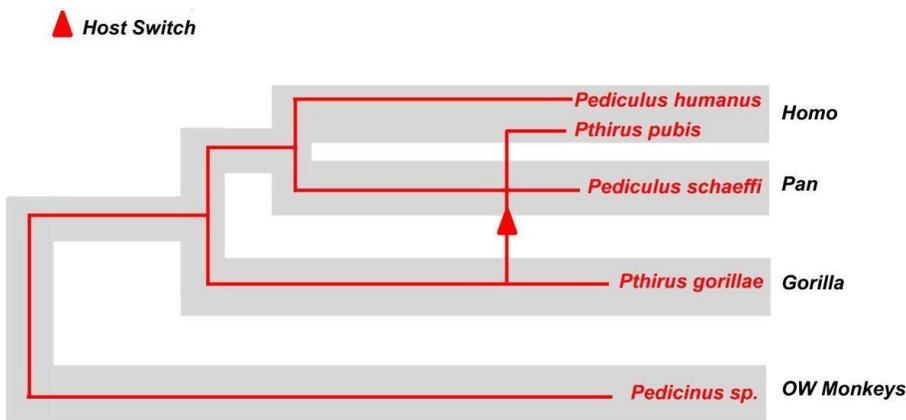

Fig 6. Reconstruction showing perfect cospeciation between with hosts and parasites with the exception of a single host switch of *Pthirus sp.* from gorillas to humans (marked by an arrow). Image from Reed et al, 2007.

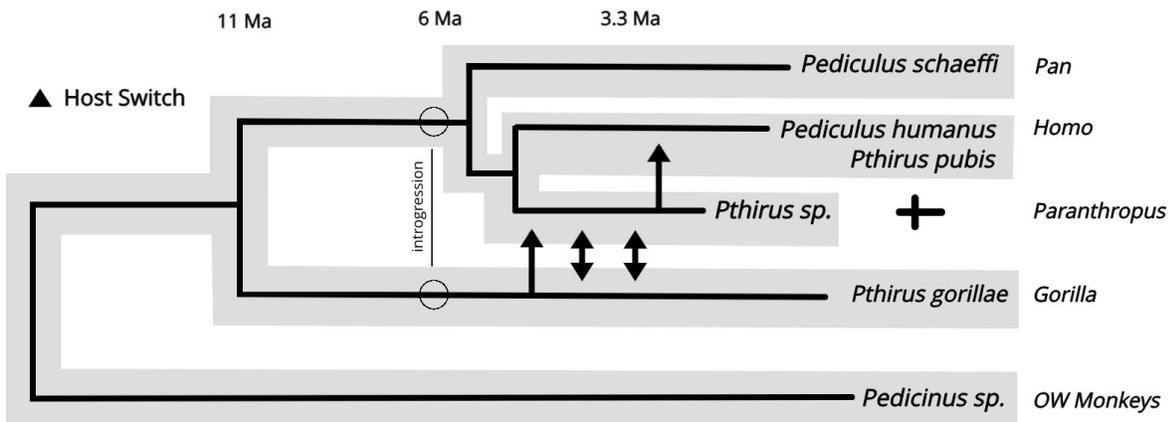

Fig 7. Coevolutionary reconstruction of primate lice and their hosts, with *Paranthropus* as an intermediary in the *Pthirus* host switch, a scenario where *Paranthropus* and *Gorilla* continued to mate up to 3.3 Ma.